# Tunneling splitting of magnetic levels in Fe8 detected by $^1$H NMR cross relaxation

Y. Furukawa, K. Aizawa, and K. Kumagai
*Division of Physics, Graduate School of Sciences, Hokkaido University, Sapporo 060-0810, Japan*

R. Ullu and A. Lascialfari
*Dipartimento di Fisica "A. Volta" e Unità INFM di Pavia, 27100 Pavia, Italy*

F. Borsa[a]
*Dipartimento di Fisica "A. Volta" e Unità INFM di Pavia, 27100 Pavia, Italy and Department of Physics and Astronomy, Iowa State University and Ames Laboratory, Ames, Iowa 50011*

(Presented on 14 November 2002)

Measurements of proton NMR and the spin lattice relaxation rate $1/T_1$ in the octanuclear iron (III) cluster $[\text{Fe}_8(\text{N}_3\text{C}_6\text{H}_{15})_6\text{O}_2(\text{OH})_{12}] \cdot [\text{Br}_8 \cdot 9\text{H}_2\text{O}]$, in short Fe8, have been performed at 1.5 K in a powder sample aligned along the main anisotropy $z$ axis, as a function of a transverse magnetic field (i.e., perpendicular to the main easy axis $z$). A big enhancement of $1/T_1$ is observed over a wide range of fields (2.5–5 T), which can be attributed to the tunneling dynamics; in fact, when the tunneling splitting of the pairwise degenerate $m = \pm 10$ states of the Fe8 molecule becomes equal to the proton Larmor frequency a very effective spin lattice relaxation channel for the nuclei is opened. The experimental results are explained satisfactorily by considering the distribution of tunneling splitting resulting from the distribution of the angles in the hard $xy$ plane for the aligned powder, and the results of the direct diagonalization of the model Hamiltonian. © *2003 American Institute of Physics.* [DOI: 10.1063/1.1540052]

The study of quantum tunneling of magnetization (QMT) has received a new impulse following the discovery of this phenomenon in magnetic molecular clusters.[1] A system that has been widely investigated is $[\text{Fe}_8(\text{N}_3\text{C}_6\text{H}_{15})_6\text{O}_2(\text{OH})_{12}] \cdot [\text{Br}_8 \cdot 9\text{H}_2\text{O}]$, referred to hereafter as Fe8.[2] QMT in Fe8 is associated with the existence in the high spin ground state ($S=10$) of pairwise degenerate $\pm m$ magnetic levels separated by an energy barrier (about 27 K) due to easy axis crystal field anisotropy in the $z$ direction. The occurrence of QMT is related to the splitting of the magnetic levels by an amount $\Delta_T$ which is due to off diagonal terms in the magnetic Hamiltonian arising from anisotropy in the $xy$ plane, intermolecular dipolar interactions, and hyperfine interactions. The very small tunneling splitting in the ground state of Fe8 has been measured in the millikelvin temperature range by the Landau-Zener method.[3] At higher temperature the tunnel splitting $\Delta_T$, particularly for the low lying $m$ magnetic states, is much smaller than the level broadening and thus difficult to detect. However, by applying a magnetic field perpendicular to the easy axis one can increase $\Delta_T$ of all levels while leaving the symmetry of the double well potential intact.[3,4] This circumstance has allowed the determination of large tunneling splitting directly by ac susceptibility measurements[5] and indirectly by specific heat measurements.[6] It also raises the possibility of detecting the tunnel splitting by microscopic spectroscopic means such as NMR. In fact the coherent or incoherent tunneling between pairwise degenerate states across the anisotropy barrier generates a fluctuation of the hyperfine field at the nuclear site centered at a frequency $\omega_T = h\Delta_T$. If the external magnetic field is tuned so that the nuclear Larmor frequency $\omega_L = \omega_T$ one expects an enhancement of the nuclear spin lattice relaxation rate $1/T_1$ due to the very effective relaxation channel opened by the transfer of energy from the nuclear Zeeman reservoir to the tunneling reservoir. This effect, which has been widely used to study the tunneling dynamics of atoms and molecular groups,[7] is used here to detect the tunneling splitting of the $m = \pm 10$ ground state in Fe8. It is noted that the main difference between the NMR detection of tunneling splitting of magnetic levels in nanomagnets and of rotational or vibrational levels in molecular groups is that in the former case the applied transverse field increases both the tunneling splitting and the Zeeman splitting at the same time. Thus when cross relaxation occurs we measure the splitting due primarily to the transverse field.

Measurements of the proton spin lattice relaxation time $T_1$ have been performed at 1.5 K in an oriented powder of Fe8 as a function of an external field applied perpendicular to the orientation axis $z$ of the powder, which coincides with the main easy magnetic axis of the molecule. Proton NMR in oriented powders of Fe8 has been reported previously,[8] and we refer to that paper for details about the experimental set up. Upon decreasing the temperature the NMR spectrum becomes broad and with a structure which reflects the different local fields at the different proton sites in the molecule. By measuring the proton spin lattice relaxation rate $1/T_1$ (NSLR) at any of the peaks one obtains information about the fluctuations of the magnetization $M$ of the molecule. The measurements reported here were obtained on two peaks labeled as $P_1$ and $P_2$ both shifted with respect to the proton

---

[a]Electronic mail: borsa@ameslab.gov

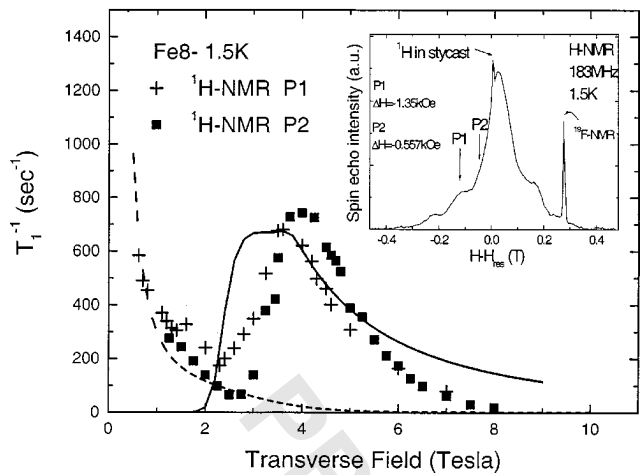

FIG. 1. Proton spin lattice relaxation rate in oriented powder of Fe8 plotted versus applied magnetic field nominally in the *xy* plane. The two sets of data refer to measurements done on different lines present in the NMR spectrum at 1.5 K. The dashed and full curves are theoretical curves representing the thermal fluctuation contribution and the tunneling contribution to $1/T_1$, respectively (see text). The data are affected by an uncertainty of about 10%. The error bars were omitted for clarity.

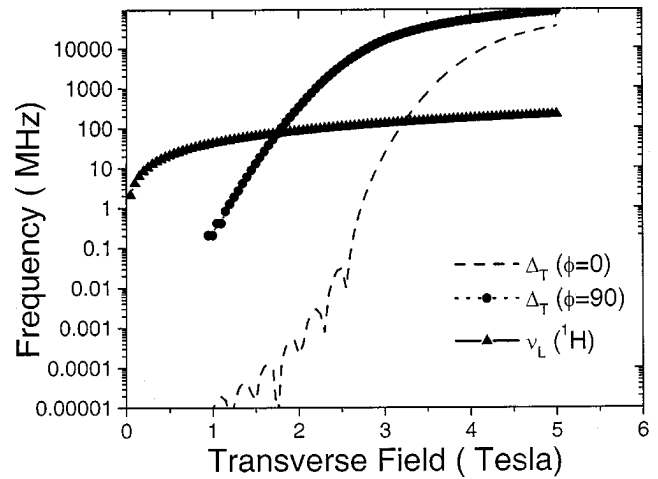

FIG. 2. Calculated tunneling splitting in frequency units as a function of the applied field perpendicular to the *z* axis for a Fe8 molecule. The two curves refer to two limiting angles formed by the field *H* with the *x* hard axis in the *xy* plane. The triangles represent the proton Larmor frequency corresponding to the applied field *H*.

Larmor frequency.[8] The recovery of the nuclear magnetization following a saturation sequence of radio frequency pulses was found to be nonexponential. The relaxation rate parameter $1/T_1$ was extracted from the slope of the initial part of the recovery curve which approximates the tangent at the origin, a parameter representing the weighted average of the relaxation rate for the different protons in the molecule and the different orientations of the grains in the *xy* plane with respect to the external transverse field.[8] The results are shown in Fig. 1. The initial decrease of $1/T_1$ on increasing the transverse field up to about 2 T can be explained in terms of the thermal fluctuations of the magnetization *M* of the molecule due to the spin-phonon induced transitions among the magnetic *m* sublevels. The dashed line in Fig. 1 represents the result of a theoretical calculation based on a simple model of spin lattice relaxation described in Refs. 8 and 9. We used the same spin-phonon coupling constant as in Ref. 8, while the hyperfine coupling constant used here is bigger since the results in Fig. 1 refer to NMR lines shifted with respect to the Larmor frequency and thus experiencing a larger hyperfine interaction. The magnetic energy levels that enter in the formula[8] for $1/T_1$ are the ones calculated from the exact diagonalization of the model Hamiltonian in a transverse field as discussed below. The broad peak in $1/T_1$ extending from 2 up to 5 T in Fig. 1 is the new feature which we will analyze in the following in terms of tunneling dynamics of the molecular magnetization.

The tunneling splitting of the $m=\pm 10$ ground state in the presence of a transverse field can be calculated from the diagonalization of the model Hamiltonian

$$H = DS_z^2 + E(S_x^2 - S_y^2) - g\mu_B \mathbf{S}\cdot\mathbf{H}, \qquad (1)$$

where $D<0$ defines *z* as the main easy axis anisotropy and $E>0$ is the in plane anisotropy which defines *y* as a secondary easy axis and *x* as the hard axis. The crystal field parameters *D* and *E* can be determined very accurately by high frequency electron paramagnetic resonance (EPR) experiments.[10,11] When $\mathbf{H}\perp \mathbf{z}$ the diagonalization of Hamiltonian (1) leads to energy levels that change as a function of the field and cannot be described simply in terms of the *m* quantum number. In particular, the initial degenerate $\pm m$ levels undergo a splitting which increases with increasing transverse field and that we call tunnel splitting $\Delta_t$. Its value in frequency units for $m=\pm 10$ calculated with $D=-0.295$ K and $E=0.046$ K is shown in Fig. 2 for two limiting orientations of the transverse field in the *xy* hard plane. In the same Fig. 2 we plot the field dependence of the proton Larmor frequency to put in evidence that the latter does intercept the tunneling frequency at fields ranging from 1.75 to 3.25 T depending on the orientation of **H** in the *xy* plane. In the presence of tunneling of the magnetization between the two lowest magnetic states the hyperfine field at the proton site fluctuates and can generate nuclear spin lattice relaxation via a weak collision mechanism, whereby $1/T_1$ is proportional to the spectral density of the fluctuations at the Larmor frequency: $1/T_1 \propto J(\omega_L)$. In order to describe the nuclear relaxation process associated with the tunneling dynamics of the magnetization we assume an expression similar to the one derived for the relaxation due to tunneling dynamics of particles and/or of molecular groups such as $CH_3$:[7,12]

$$1/T_1 = A(\Gamma/\{\Gamma^2 + [\Delta_T(H)/h - \gamma_N H]^2\} + \Gamma/\{\Gamma^2 + [\Delta_T(H)/h + \gamma_N H]^2\}) \qquad (2)$$

where *A* is the square average of an effective time dependent hyperfine coupling constant, $\gamma_N$ is the nuclear gyromagnetic ratio, $\Delta_T(H)$ is the field dependent tunneling splitting, and $\Gamma$ is an effective linewidth parameter which incorporates the effect of the broadening of the molecular magnetic levels as well as the broadening of the NMR line. As mentioned above, the measured relaxation rate is an average one. Two averages have to be considered. First, for a given proton NMR line (see Fig. 1) one has to average over the different hyperfine constants $A$ which couple the different protons irradiated with the Fe magnetic moments. This average can be incorporated in the constant $A$ which will be treated as a fitting parameter. The second average is due to the fact that the oriented powder contains grains with different orientations of the transverse field $H$ in the $xy$ plane. By defining as $\phi$ the angle between the transverse field $H$ and the hard axis $x$, one can calculate the average relaxation rate as [the second term in Eq. (2) can be neglected in our case]

$$1/T_1 = A \int_0^{\pi/2} N(\vartheta_L)\Gamma/\{\Gamma^2 + [\Delta_T(H,\phi)/h - \gamma_N H]^2\}d\phi \quad (3)$$

where $\Delta_T(H,\phi)$ is the tunneling splitting calculated by diagonalizing Hamiltonian (1) and shown in Fig. 2 for two limiting angles only. In performing the average we assumed a uniform distribution of grains regarding the orientation of the transverse field $H$ in the $xy$ plane. On the other hand, an important role in the average (3) is played by the distribution of grain orientation with respect to the angle $\vartheta$ between the field $H$ and the $z$ axis. In fact, although the powder is oriented, there is a distribution of misalignment angles around the ideal condition $\vartheta = 90$. On the basis of magnetization measurements we estimate a distribution of angles which can be described by a Gaussian function with mean square deviation of 3° around the ideal angle of 90°. For the grains that are not perfectly aligned there is a longitudinal component of the field $H_L = H_{ext} \cos(\vartheta)$, which removes the pairwise degeneracy for the $\pm m$ states, thus preventing the tunneling effect. In order to take into account this effect we performed the integral (3) with the condition that the number density $N(\vartheta_L)$ is given by the number of grains whose angle $\vartheta$ is included between 90° and a limiting angle $\vartheta_L$ defined by the condition $10 G \mu_B H \cos(\vartheta_L) \leq \Delta_T(H)$. The criterion chosen here is that the shift of the $m$ states due to the longitudinal component of the applied field $H_{ext}$ must be less than the tunneling splitting in order to maintain a tunneling dynamics. The result of the calculation according to Eq. (3) with the above assumptions is shown in Fig. 1. The coupling constant $A$ is a fitting parameter which rescales the amplitude of the peak. The other fitting parameter is the width $\Gamma$ of the Lorenzian function in Eqs. (2) and (3) which was chosen to be 1200 MHz in the theoretical curve in Fig. 1. Although the agreement is only qualitative it should be considered a success in view of the crudeness of the model and of some of the assumptions made. Since the degree of alignment of the powder appears to be crucial in our model we have performed measurements of the proton $T_1$ in the same aligned powder sample by introducing a tilt angle $\varphi = 90° - \vartheta$ between the applied field $H_{ext}$ and the $xy$ plane. This should result in a decrease of the maximum of $1/T_1$, as indeed is found experimentally (Fig. 3). The full lines in Fig. 3 represent the theoretical calculation of the integral Eq. (3), when all parameters are kept the same except for the tilt angle and the transverse field which is taken to be $H = H_{ext} \cos(\varphi)$. The theoretical curves appear to reproduce correctly the trend of the experimental data, thus confirming the basic validity of the model adopted. In summary, we have demonstrated that one can detect the tunneling splitting of the magnetic ground state in Fe8 by applying a transverse field and looking for the enhancement of the proton $1/T_1$ when the tunneling frequency matches the nuclear Larmor frequency. The interpretation of the results yields only a semiquantitative agreement in view of the averaging effects involved in the measurements in oriented powder samples. More quantitative and direct information regarding the tunneling splitting and the broadening of the molecular magnetic levels should be possible by performing measurements in single crystals, measurements which are currently under way.

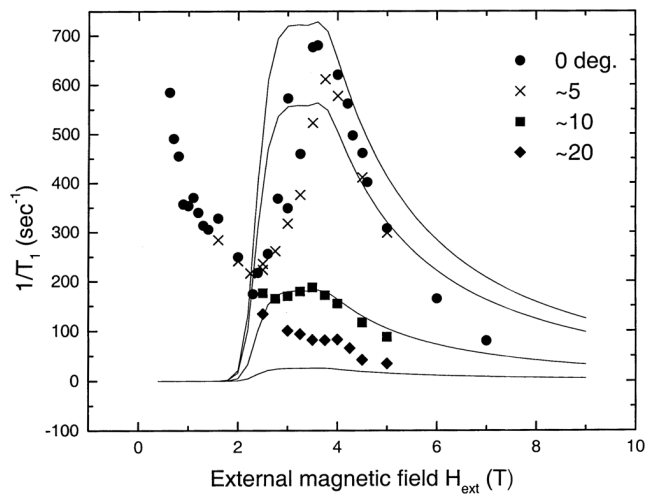

FIG. 3. Proton spin lattice relaxation rate in oriented powder of Fe8 for different angles of the external field with respect to the $xy$ plane. The full lines are the theoretical predictions for the tunneling contribution to $1/T_1$ in order of decreasing intensity for increasing angle. The data are affected by an uncertainty of about 10%. The error bars were omitted for clarity.

This work was supported by a grant-in-aid for Scientific Research from the Ministry of Education, Science, Sport and Culture of Japan. The work in Pavia was part of the INFM program PRA-MESMAG. Ames Laboratory is operated for the U.S. Department of Energy by Iowa State University under Contract No. W-7405-Eng-82. The work at Ames Laboratory was supported by the Director of Energy Research, Office of Basic Energy Sciences.